\documentclass[aps,prl,twocolumn,groupedaddress,showpacs]{revtex4}

\usepackage{graphicx}
\usepackage{dcolumn}
\usepackage{bm}
\usepackage{hyperref}

\begin{document}


\title{Bose-Einstein Condensation and Spin Mixtures of Optically Trapped Metastable Helium}


\author{G.~B.~Partridge,  J.-C.~Jaskula, M. Bonneau, D.~Boiron, C.~I.~Westbrook}
\affiliation{Laboratoire Charles Fabry de l'Institut d'Optique, Universit\'{e} Paris Sud, CNRS,
Campus Polytechnique RD 128 91127 Palaiseau France\\}


\date{\today}

\begin{abstract}
We report the realization of a BEC of metastable helium-4 atoms ($^4$He$^*$) in an all optical potential.  Up to 10$^5$ spin polarized $^4$He$^*$ atoms are condensed in an optical dipole trap formed from a single, focused, vertically propagating far off-resonance laser beam.
 The vertical trap geometry is chosen to best match the resolution characteristics of a delay-line anode micro-channel plate detector capable of  registering single He$^*$ atoms.
  We also confirm the instability of certain spin state combinations of $^4$He$^*$ to two-body inelastic processes, which necessarily affects the scope of future experiments using optically trapped spin mixtures. In order to better quantify this constraint, we measure spin state resolved two-body inelastic loss rate coefficients in the optical trap.
\end{abstract}


\maketitle


When a helium atom is placed in the long lived 2$^3$S$_1$ metastable state, it behaves in many ways like any number of other atomic species used in trapping and cooling experiments.  In particular, metastable helium (He$^*$) can be manipulated by common laser and evaporative cooling techniques, the application of the which have culminated in the Bose-Einstein condensation of $^4$He$^*$ \cite{A.Robert04202001,PhysRevLett.86.3459,PhysRevA.73.031603,Dall2007255} and the realization of a degenerate Fermi gas of $^3$He$^*$ \cite{PhysRevLett.97.080404}.   Despite many similarities however, ultracold metastable helium also presents unique rewards and challenges in relation to the more traditional ultracold alkali atomic systems due to the almost 20 eV of potential energy per atom that is stored in the metastable excited state.

One of the most appealing aspects of He$^*$ is the range of possibilities it offers for the detection of individual atoms, as well as for the real-time monitoring of the trapped gas. Through the use of a micro-channel plate (MCP) and delay-line anode, it is possible to measure 3-dimensional (3D) time of flight (TOF) distributions with single atom resolution \cite{M.Schellekens10282005}.  In addition, the stream of ions produced by the collision induced ionization of trapped atoms provides a realtime $\textit{in-situ}$ readout of the density of the trapped gas.  Using these techniques, we have previously studied phenomena such as the onset of a BEC in real-time \cite{Seidelin1464-4266-5-2-367}, the Hanbury Brown-Twiss effect for bosonic and fermionic atoms \cite{M.Schellekens10282005,Jeltes:HBTNature}, and the degenerate 4-wave mixing of matter waves resulting from collisions between condensates \cite{PhysRevLett.99.150405,krachmalnicoff-2009}.  The pairing of an MCP and a phosphor screen has also allowed the direct imaging of He$^*$ atom laser modes \cite{Dall:07,PhysRevA.79.011601,PhysRevA.81.011602}.

The large energy stored in the metastable state that enables the exceptional detection schemes used in these experiments comes at a price, however: the loss of atoms in the trapped gas through the enhancement of collisional ionization processes \cite{PhysRevLett.59.2279,Julienne:89}.  Indeed, were it not for the fact that such losses are strongly suppressed in a polarized gas \cite{Julienne:89,PhysRevLett.73.3247,PhysRevA.61.050702}, it would not have been possible to reach the atomic densities sufficient for condensation.

To date, experiments involving degenerate He$^*$ have been largely based upon magnetic trapping, if for no other reason than the magnetic trap guarantees the spin polarization (for the case of $^4$He$^*$) that enables the high densities that are necessary to reach degeneracy.  In recent years however, the power and flexibility of all optical potentials formed from far-detuned laser beams, together with the lifting of the constraint of spin polarization, has facilitated access to diverse fields of study including models of condensed matter systems and spinor physics \cite{RevModPhys.80.885}.  The combination of custom engineered trapping potentials, the spin degree of freedom, and the power of single atom detection of He$^*$ should open many exciting possibilities.

As a first step in this direction, we have constructed an optical dipole trap for He$^*$.  In this paper, we describe the initial implementation of this trap and the subsequent production of a BEC.  Moreover, we report measurements of the dominant ionizing loss rate constants between different spin states, the values of which will determine the feasibility of future studies involving non-polarized gases.

\section{\label{sec:bec}BEC of $^4$He$^*$ in an Optical Trap}
Up to $N$ = 10$^8$ $^4$He$^*$ atoms at a temperature of $T$ $\sim$ 300 $\mu$K are initially loaded from a magneto-optical trap
into a Ioffe-Pritchard type magnetic trap in the 2$^3$S$_1$, m$_J$ = 1 magnetic substate where they are cooled to $T$ = 150 $\mu$K by a retro-reflected 1D Doppler beam \cite{Schmidt:03}.  An RF forced evaporation ramp is then applied for $\sim$ 4 s to further cool the gas to $T$ = 15 $\mu$K, $N \sim$ $5\cdot10^6$, after which time the optical trap, formed from a single far detuned 1.5 $\mu$m wavelength laser beam focussed to a $1/e^2$ gaussian beam waist radius of $w_o$ = 43 $\mu$m, is ramped up to a full power of $P$ $\sim$ 1.5 W over approximately 1 s.  The magnetic trap is then switched off in a time on the order of 100 $\mu$s, and a bias magnetic field $B_{o}$ = 4 G is applied in the horizontal direction in order to maintain the polarization of the atoms.

At full power, the trap provides  $U_o$ $\sim$ 50 $\mu$K of confinement in the radial (horizontal) direction, though due to gravity, an effective ``lip'' is formed at the bottom of the axial (vertical) potential.  The height of this lip and correspondingly the effective trap depth, $U_{eff}$, is strongly dependent on the optical trap laser power as well as on residual vertical magnetic field gradients.  Under typical conditions and at full optical power, the trap depth is reduced in this direction by a factor of three.  To further characterize the trapping potential, trap oscillation frequencies are measured via the resonant loss of the atoms resulting from parametric excitations driven by intensity modulation of the laser.  At full trap depth, the approximated harmonic trap oscillation frequencies are $\omega_r / 2\pi$ = 2.4 kHz and $\omega_z / 2\pi$ = 15 Hz.
At this trap depth, we also measure a background pressure limited 1/e atom lifetime of $\tau$ $\sim$ 25 s using a dilute polarized $m_J$ = 1 gas, for which two and three-body losses are highly suppressed.

\begin{figure}
 \includegraphics[width=3.5in]{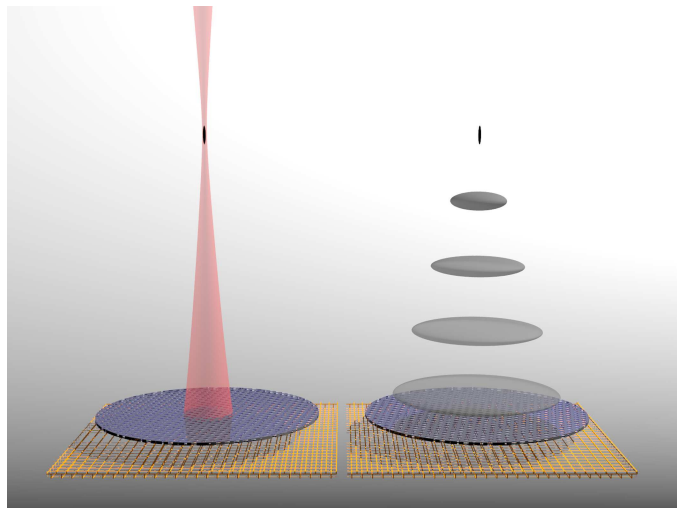}
 \caption{\label{fig:app} Trap and detector layout. The optical trap is formed from a vertically propagating red-detuned beam.  The anisotropy of the potential results in a vertically elongated condensate with typical radial and axial (vertical) Thomas-Fermi radii of $R_r$ $\sim$ 2.5 $\mu$m, and $R_z$ $\sim$ 600 $\mu$m, where $R^2_i$ = $2\mu_{3D}/m\omega^2_i$ (the 3D chemical potential $\mu_{3D}$ is defined in the text, and trap frequencies $\omega_i$ corresponding to a BEC are given in the following figure caption). After release from the trap, during the 300 ms TOF, the condensate expands into a horizontally oriented ``pancake'' distribution since the radial size increases by a factor of $\sim 3000$, while the axial size increases by only a few percent.  After their TOF, individual atoms are detected in the horizontal plane with a crossed delay line anode positioned beneath an MCP.  Combined with the time of arrival, this allows a 3D reconstruction of the TOF atomic distribution. }
 \end{figure}

After transfer from the magnetic trap, approximately $N$ = $10^6$ atoms remain in the optical trap at a temperature of $T$ = 5 $\mu$K.  After 1 s of thermalization, $N$ is reduced by 10$\%$ and the temperature reaches a steady state value of 3 $\mu$K.   Further cooling is achieved by evaporation as the trap laser intensity, $I$, is decreased from maximum, $I_o$, towards a constant non-zero value, $I_F$, in an exponential fashion over $\sim$ 4 s as $I(t) = (I_o-I_F)e^{(-t/\tau)}+I_F$, where the time constant $\tau = 1$ s.
After evaporation, the trapping beam is switched off and the atoms fall onto the MCP detector, located 47 cm below the trapping region (TOF $\sim$ 300 ms), as illustrated in fig. \ref{fig:app}.

\begin{figure}
 \includegraphics[  width=3.5in]{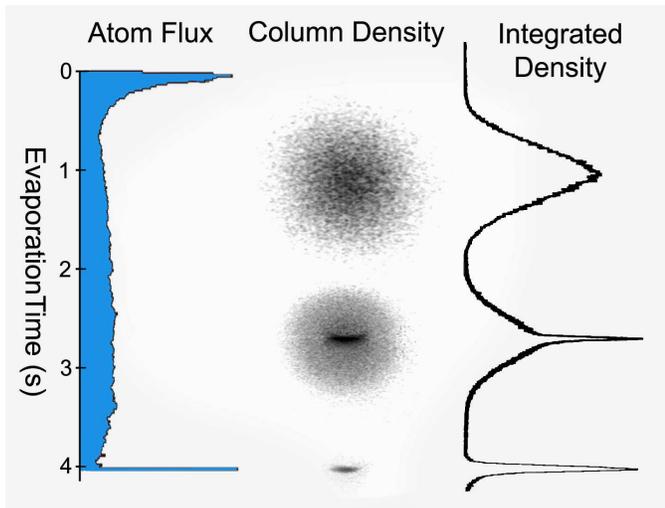}
 \caption{\label{fig:bec} Evaporation to BEC.  On the left is the flux of atoms exiting the trap over the vertical potential lip during evaporation.  These atoms remain confined by the trapping beam in the horizontal plane, and so are guided onto the MCP.  The peak near time = 0 s is associated with atoms lost during transfer from the magnetic trap and the subsequent rapid thermalization of the remaining atoms.  The peak near 4 s is the TOF density distribution of the cold atoms released at the end of evaporation.  In the center are 2D column densities obtained by integration along one horizontal axis of the 3D TOF density distributions of released clouds of various temperatures.  On the right are the corresponding axial profiles made by integration along the remaining horizontal axis.  At full trap depth, the temperature $T$ = 3 $\mu$K $>$ $T_c$, where the critical temperature for condensation $T_c$ $\approx$ 2 $\mu$K, and so the momentum distribution is thermal (top).  As the temperature is lowered below $T_c$, the distribution becomes a bimodal superposition of thermal and condensed atoms, marked by the high visibility of the pancake shaped condensed fraction contrasted with the uniformly distributed thermal component (middle).  When the trap intensity is lowered until $I$ $\approx$ 0.5 $I_o$, the temperature $T$ $\ll$ $T_c$, and the thermal component is removed, leaving a quasi-pure condensate of up to $N$ = 10$^5$ atoms with condensed atom fraction $N_c/N$ $\geq$ 0.9 (bottom).  At this trap depth ($U_{eff}/k_B$ $\sim$ 0.25 $\mu$K), the trap frequencies are $\omega_r/2\pi$ = 1.5 kHz and $\omega_z / 2\pi$ = 6.5 Hz.}
 \end{figure}

The final degeneracy of the atomic cloud can be controlled by either varying the offset of the evaporation trajectory, $I_F$, or by interrupting the trajectory before the final trap depth is reached.  Figure \ref{fig:bec} shows time of flight momentum distributions corresponding to three different final trap depths.   The distribution of the condensed fraction exemplifies the striking inversion of aspect ratio of the initially cigar shaped condensate that results from the anisotropic expansion during the long TOF.

Due to the vertical orientation of the beam relative to gravity, the escaping atoms exit the trap primarily over the lip at the bottom of the trap, along the propagation direction of the laser, during evaporation.  This results in an intense beam of optically guided atoms that is registered on the MCP directly below the trap.  This signal of evaporated atom flux vs. time, shown in fig. \ref{fig:bec}, has proven useful in the optimization of atom transfer from the magnetic trap and subsequent evaporation, and provides a unique realtime method of monitoring the evolution of the cooling gas.  Moreover, the same mechanism has recently been used to guide an atom laser formed from a He$^*$ BEC  \cite{PhysRevA.81.011602}.

The primary motivation for the vertical orientation of this trap, however, is to create a better ``mode matching'' between the time of flight density distribution of the condensate and the resolution capabilities of the MCP detector, for which the resolution in the vertical direction, $\sigma_z \approx 3$ nm, is much higher than that of the horizontal direction,  $\sigma_{x,y} \approx 300 \mu$m.  In addition to aligning the small dimension of the ``pancake'' BEC  with the high resolution direction of the detector (see fig. \ref{fig:app}), this geometry is expected to improve the contrast of measurements of Hanbury Brown-Twiss type correlations, since the shortest correlation length, associated with the vertically elongated potential, is now aligned with the vertical axis \cite{M.Schellekens10282005,Jeltes:HBTNature}.

 Finally, we note that condensates formed at the lowest trap depths, such as those shown in fig. \ref{fig:bec}, approach the crossover to 1D that occurs when the 3D chemical potential falls below the energy of the first harmonic oscillator state of the radial potential \cite{PhysRevLett.87.080403,PhysRevLett.87.130402}:  $\mu_{3D} = \frac{1}{2} \hbar \bar{\omega} (15 N a/\bar{a})^{2/5}  < \hbar \omega_r$, where $\bar{\omega} = (\omega_r^2 \omega_z)^{1/3}$ is the geometric mean of the trap frequencies, $\bar{a} = \sqrt{\hbar/(m\bar{\omega})}$ is the characteristic oscillator length, $a$ = 7.5 nm is the s-wave scattering length \cite{PhysRevLett.96.023203}, and $m$ is the $^4$He atomic mass.  Condensates with further reduced number $N$ $\sim 10^4$ remain well within our detection capabilities, and can in fact cross this boundary.
\section{\label{sec:losses}Inelastic Losses in a Thermal Spin Mixture of $^4$He$^*$}

As already mentioned, it is primarily the stored internal energy of the metastable $^4$He$^*$ atom that distinguishes it from more commonly trapped alkali atoms.  This characteristic allows single atom detection, but also leads to increased susceptibility to inelastic loss processes whose rates are enhanced by Penning and associative ionization:
\begin{equation}
\text{He}^* + \text{He}^* \rightarrow \left\{ \begin{array}{l} \text{He} +\text{He}^+ +e^- \\ \text{He}_2^+ +\text{e}^- \end{array}
\right.
\label{eq:ionizingprocesses}
\end{equation}

In an external field, the 2$^3$S$_1$ state of $^4$He$^*$ has three magnetic substates m$_J$ = $\pm 1$ and m$_J$ = 0.   In a magnetic trap, the required polarization of the trapped gas in the m$_J$ = 1 state suppresses these processes.  In an optical trap, however, atoms in all spin projections can be confined.  To better comprehend the feasibility of exploiting this added degree of freedom, we now turn to state-resolved measurements of inelastic loss from spin mixtures of thermal atoms in the optical trap.

At low temperatures such as those considered in this work, where s-wave atomic collisions dominate, inelastic collisions between $^4$He$^*$ atoms can take place through the $^5\Sigma^+_g$ potential, for colliding atoms with total spin $M$ $\neq$ 0, or the $^1\Sigma^+_g$ potential for $M$ = 0, where $M$ = $\Sigma m_J$ is the sum of the spin projections of the colliding atoms. Ionization processes are spin-forbidden for collisions through the former, which is the origin of the suppression of inelastic losses in spin polarized samples.  Moreover, any other combination of input states with $M$ $\neq$ 0 should be likewise suppressed.  The collisions that remain are those between atoms with $M$ = 0, namely collisions involving two $m_J$ = 0 atoms or one each of $m_J$ = 1 and $m_J$ = -1 \cite{Julienne:89,PhysRevA.61.060703,PhysRevA.64.042710}.

\begin{figure}
 \includegraphics[bb=14 14 352 249, width=3.45in]{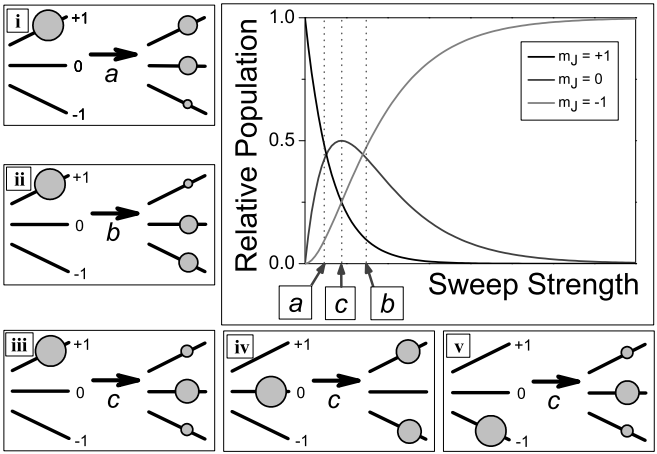}
 \caption{\label{fig:rf} RF sweep schemes.  The large panel shows the solution to the three level L-Z system for input state $m_J$ = 1, plotted vs. ``Sweep Strength'' which is proportional to the RF power for a sufficiently broad sweep of fixed duration.  Panels [i-v] show the effects of ramps of various strengths (labeled ``a'', ``b'', and ``c'') on different input states.  Note that the total population is an incoherent mixture of spin states, so only single state inputs are considered.  Panels [i-iii] correspond directly to the main panel, and [v] is equivalent to [iii] with the $m_J$ = $\pm$1 states reversed.  The behavior of panel [iv] can not be inferred from the main panel, since the input state in this case is $m_J$ = 0, though we have confirmed the result both experimentally and numerically, as described in the text. }
 \end{figure}

By creating mixtures of various spin states and monitoring their populations as a function of time, we can test this prediction and also quantitatively measure the loss rates for the dominant processes.
As previously described, atoms are initially loaded into the optical trap and cooled in the 2$^3$S$_1$, m$_J$ = 1 magnetic substate, the same as is trapped in the magnetic trap.  After transfer, the optical trap depth is lowered such that the gas temperature approaches, but does not go below, $T_c$.  This cooling phase is followed by an adiabatic recompression to full intensity, where the trap is held for $\sim$ 1 s to ensure thermalization.  Typical values following this trap trajectory are, $N = 5\times10^5$, and $T$ = 2.0 $\mu$K, so that $T/T_c$ $\gtrsim$ 1.25 with peak density $n_0$ $\approx$ 2.5 $\times$ 10$^{12}$ cm$^{-3}$.

Following cooling and recompression, spin mixtures are created from the polarized $m_J = 1$ gas by a non-adiabatic RF sweep across the transition between adjacent levels, $m_J = \pm1 \leftrightarrow m_J = 0 $.  For the current work, we use 2 ms linear sweeps of 3 MHz centered at $\sim$ 11 MHz at three separate power settings to prepare the atom spin states in different proportions.  Figure \ref{fig:rf} shows a numeric solution to a three level Landau-Zener (L-Z) problem as a function of sweep strength, which is proportional to the RF power divided by the linear sweep rate \cite{PhysRevLett.78.582}.  As illustrated in fig. \ref{fig:rf}, for the case of the weak sweep ``a'', the final population is approximately evenly split between the $m_J$ = +1 and $m_J$ = 0 states, with a small amount ($\sim$ 10\%) of $m_J$ = -1.  A sweep of intermediate power, ``c,'' places half the atoms in the $m_J$ = 0 state, with the remaining half split evenly between the $m_J$ = $\pm$ 1 states.  As the RF power is increased further, the sweep ``b'' splits the population between $m_J$ = 0 and $m_J$ = -1, with a small amount ($\sim$ 10\%) of $m_J$ = 1 remaining.

Before making quantitative measurements of the loss rates, we first check the overall relative stability of various combinations of spins. After spin preparation, the trap is held at full power for a variable time, and is then suddenly switched off.   A Stern-Gerlach vertical magnetic field gradient is then applied in order to impart a spin-dependent momentum component, such that after TOF, the arrival times of the three spin state populations are different and each can be counted separately by the MCP.

 \begin{figure}
 \includegraphics[bb =18 15 301 241, width=3.35in]{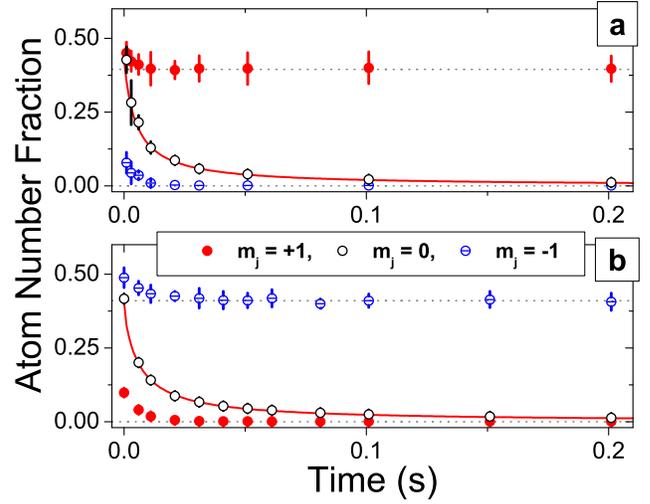}
 \caption{\label{fig:mixgraph} Loss processes in imbalanced mixtures.  Two initial spin state populations prepared using ``a'' and ``b'' of fig. \ref{fig:rf} are plotted vs. time to determine the stable and unstable spin combinations.  The error bars correspond to the standard deviation of the mean derived from repeated measurements.  The solid curve through the $m_J$ = 0 points is a fit to a two-body loss process corresponding to a preliminary value $\beta^{prelim}_{00}$ $\approx$ 7 $\times$ 10$^{-10}$ cm$^3$/s, and is shown for reference, since in this case, systematic uncertainties in atom number and temperature that result from residual distortions associated with the Stern-Gerlach field limit the precision of the measurement. The dotted horizontal lines are shown for reference.  }
 \end{figure}

Figure \ref{fig:mixgraph}a and \ref{fig:mixgraph}b show measurements of losses in imbalanced spin mixtures prepared with sweeps ``a'' and ``b'' of fig. \ref{fig:rf}, respectively.
In fig. \ref{fig:mixgraph}a, the populations of all three spin states initially decay until the $m_J$ = -1 is depleted, at which time the $m_J$ = 1 population stabilizes while the $m_J$ = 0 population continues to decay for the entire duration.  Figure \ref{fig:mixgraph}b shows a similar measurement in which the roles of the $m_J$ = 1 and $m_J$ = -1 have been reversed.  The observed behavior confirms the instability of the ($m_J$,$m'_J$) = $(1,-1)$ and $(0,0)$ mixtures, as well as the relative stability of the ($m_J$,$m'_J$) = (0, 1), (0,-1), (1,1), and (-1,-1) mixtures.    The extent of the stabilization of the $m_J$ = 1 in \ref{fig:mixgraph}a and the  $m_J$ = -1 in \ref{fig:mixgraph}b in the presence of the remaining $m_J$ = 0 gas places a coarse upper limit on the loss rate coefficients $\beta_{10,-10,11,-1-1}$ $\lesssim$ 10$^{-13}$ cm$^3$/s, though two-body loss processes below this level are indistinguishable from the background one-body losses under the present experimental conditions.
 This finding is consistent with the suppression of ionizing collisions already observed for spin polarized $m_J$ = 1 samples, and moreover, is a direct demonstration of the extension of this suppression to other atom pairs with total spin $M$ $\neq$ 0.

 With the stable and unstable combinations of spin states identified, we now focus on a careful and quantitative treatment of the more dominant loss rates between states ($m_J$,$m'_J$) = $(0,0)$ and $(1,-1)$.  
  In the following, we model the local atomic density of the thermal gas, $n = n(\textbf{r})$, as
\begin{equation}
\frac{dn}{dt} = -\Gamma n-\beta n^2 - L_3 n^3,
\label{eq:lossdiffeqbasic}
\end{equation}
where  $\Gamma$ = $1/\tau$ = $0.04$ s$^{-1}$ is the measured one-body loss rate due to background gas collisions and off-resonance scattering of trap laser photons, $\beta$ is the two-body loss rate coefficient, and $L_3$ accounts for three-body contributions to the loss rate.
Though we keep the three-body term $L_3$ for the time being, we intentionally use low density gases in order to minimize its contribution and isolate two-body effects.

In order to compare with our data, eqn. \ref{eq:lossdiffeqbasic} is put into terms of the total number, $N$, by spatial integration over the extent of the inhomogeneous density distribution of the trapped cloud, $n(\textbf{r})$.
\begin{equation}
\frac{dN}{dt} = -\Gamma N - \beta N \langle n \rangle - L_3 N \langle n^2 \rangle ,
\label{eq:lossdiffeqmoments}
\end{equation}
 where $\langle n^q \rangle = (1/N)\int d^3r [n(\textbf{r})]^{q+1}$.

If we consider populations of interacting magnetic substates ($m_J, m'_J$) = ($i,j$) with numbers $N_{i(j)}$  in the $i^{th} (j^{th})$ substate, eqn. \ref{eq:lossdiffeqmoments} becomes
\begin{equation}
\frac{dN_{i(j)}}{dt} = -\Gamma N_{i(j)}-\beta_{ij}N_{i(j)}\langle n_{j(i)} \rangle- L_3 N_{tot}\langle n_{tot}^2 \rangle ,
\label{eq:lossdiffeqN}
\end{equation}
where $\beta_{ij}$ is the partial two-body rate coefficient for loss resulting from collisions between states ($m_J$,$m'_J$) = ($i,j$), and $\langle n_{j(i)} \rangle = (1/N_{i(j)})\int d^3r [n_i(\textbf{r}) n_j(\textbf{r})]$ is determined as in eqn. \ref{eq:lossdiffeqmoments} by spatial integration of the density distributions of both states $i, j$.
Here we have also generalized the three-body term as an effective rate constant dependent upon the sum of all spin component densities, $n_{tot}$.

\begin{figure}
 \includegraphics[bb =17 15 297 238, width=3.35in]{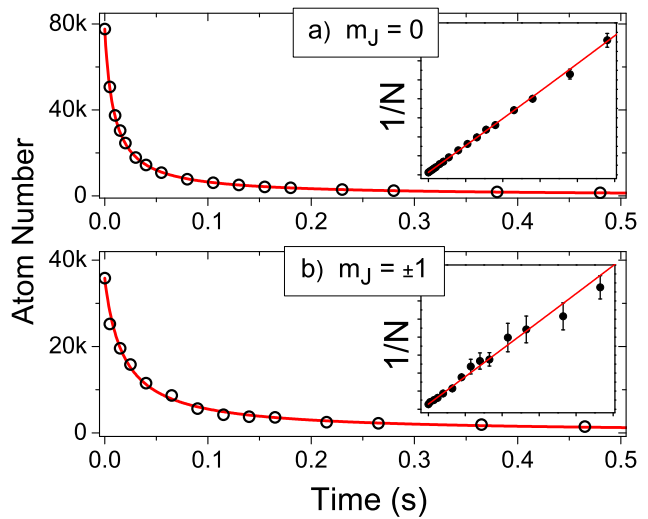}
 \caption{\label{fig:m10graph}Loss of $m_J$ = 0 and $m_J$ = $\pm$1 atoms.  The population of $m_J$ = 0 or the $m_J$ = $\pm$1 atoms are separately measured to refine measurements of the respective loss rate constants $\beta_{00}$ and $\beta_{\pm1}$.  Error bars, corresponding to the standard deviation of the mean, are equivalent to or smaller than the data point symbols (on the linear scale).  The curves show the best fit to a two-body decay process for a thermal bose gas in the true optical plus gravitational potential, including the measured one-body loss rate, as described in the text.  The insets show the same data and fits plotted as $1/N$ vs. time.  When plotted in this way, the decay resulting from purely two-body processes appears linear with time, and so the linearity of the data exemplifies the limited contribution of one and three-body loss processes.
 }
 \end{figure}

Figure \ref{fig:m10graph}a shows the loss of $m_J$ = 0 atoms from a mixture prepared with half the atoms in the $m_J$ = 0 state and the remaining half split evenly between the $m_J$ = $\pm$ 1 states as in fig. \ref{fig:rf}, panel (iii).  In order to further suppress the contribution of three-body processes in these measurements, we intentionally reduce the atom number such that the peak density is reduced to $n_0$ $\approx$ 10$^{11}$ cm$^{-3}$. Furthermore, since we have previously shown the stability of the  $m_J$ = 0 state to loss processes involving the $m_J$ = $\pm$1 states, we attribute all two-body losses in the $m_J$ = 0 to interactions among $m_J$ = 0 atoms.    To improve signal to noise on the number and temperature measurement of the $m_J$ = 0 atoms, the $m_J$ = $\pm$1 atoms that remain after the optical trap hold time are cleared by a strong transverse magnetic field gradient that is applied shortly after the optical trap is switched off (the $m_J$ = 0 atoms are not perturbed by this field gradient due to their lack of magnetic moment).

To measure the losses between $m_J$ = 1 and $m_J$ = -1 atoms, we follow a similar procedure with one added step.  Figure \ref{fig:m10graph}b shows the spin mixture that is initially prepared in the same way as in fig. \ref{fig:m10graph}a  using sweep ``c'' of fig. \ref{fig:rf}.  After the optical trap hold time, however, the RF pulse is applied a second time at the same strength before the strong clearing magnetic field gradient is applied.  We find that the second application of the RF sweep at power ``c'' splits the initial $m_J$ = 0 population into equal parts $m_J$ = $ \pm$ 1, as in fig. \ref{fig:rf}, panel (iv), which are subsequently cleared by the field gradient pulse.  Likewise, the initial $m_J$ = $ \pm$ 1 population is transferred to $m_J$ = 0 with 50$\%$ efficiency (fig. \ref{fig:rf} panels (iii), (v)), which protects it from the field gradient pulse and other stray field gradients so that it can fall to the detector unperturbed.  We have confirmed the observed behavior of the population transfers by further analysis of the three level L-Z problem \cite{Carroll:85}.    Since the populations of $m_J$ = 1 and  $m_J$ = -1 are equal, the resulting measured number $N_{\pm1}$ = $\frac{1}{2}(N_{1}+N_{-1})$ = $N_1$ = $N_{-1}$.

To find the loss rate constants, we numerically solve the loss rate eqn. \ref{eq:lossdiffeqmoments} and find the best fit to the data.
For a thermal gas in a harmonic trap, the atomic density distribution is gaussian, and can be analytically integrated to give $\langle n \rangle = N/(2\sqrt{2} V)$ and $\langle n^2 \rangle = N^2/(3\sqrt{3} V^2)$, where $V = (\omega_z / \omega_R)[m \omega_r^2/ 2 \pi k_B T]^{3/2}$ is the volume of the trapped thermal gas.

As expected for our intentionally low peak atomic densities, no contribution of the three-body term is detected in the data.  An upper limit $L_3$ $\lesssim$ 5 $\times$ 10$^{-22}$ cm$^6$/s is determined by increasing $L_3$ until its effect on the fit to the data can not be compensated, to first order, by varying the two-body term within its uncertainty.  However, even before this limit is reached, the quality of the fit is significantly degraded.  For reference, a value of $L_3$  $\approx 1 \times 10^{-26}$ cm$^6$/s has been measured in a spin polarized $m_J$ = 1 condensate of $^4$He$^*$ in a magnetic trap \cite{PhysRevLett.89.220406,PhysRevLett.86.3459,PhysRevA.73.031603}.

By varying the two-body rate constants we find preliminary values of $\beta^g_{00}$ = 7.6(4) $\times$ 10$^{-10}$ cm$^{3}$/s and $\beta^g_{\pm1}$ = 8.4(10) $\times$ 10$^{-10}$ cm$^{3}$/s, where we define $\beta^g_{00}$ as the number loss rate coefficient for $m_J$ = 0 atoms,  $\beta^g_{\pm1}$ = $\beta^g_{1,-1}$ = $\beta^g_{-1,1}$ as the number loss rate coefficient for either $m_J$ = 1 or $m_J$ = -1 atoms, and where the superscript $g$ denotes the use of the gaussian density approximation.  For the time being, the uncertainty reflects only the statistical and fitting uncertainty.

We refine the fit values of $\beta$ by integrating eqn. \ref{eq:lossdiffeqmoments} over the atomic density distribution of a bose gas at the measured temperature in the true optical plus gravitational potential.  We find that the inclusion of the slight truncation of the density distribution by the finite trap depth \cite{PhysRevA.53.381} and additional terms of the thermal bosonic density distribution result in a slight reduction of the final extracted values of $\beta_{00}$ = 6.6(4) $\times$ 10$^{-10}$ cm$^{3}$/s and $\beta_{\pm1}$ = 7.4(10) $\times$ 10$^{-10}$ cm$^{3}$/s.

In addition to the statistical errors given above, several sources of systematic error must also be included.  The dominant contributions come from the determination of the absolute number of trapped atoms and the characterization of the trapping potential.  Together with residual uncertainties in temperature \footnote{We compensate, to first order, for a small ($\sim$ 10$\%$) systematic increase in temperature that occurs as the gas decays by rescaling the measured number to account for the resultant slight decrease in peak atomic density.} these combine to give an overall estimate of $\frac{1}{\beta}\delta\beta_{sys}$ $\approx$ $25\%$.

For a comparison of relative rates, the systematic errors are largely reduced.  We find that within the remaining uncertainty that $\beta_{00} = \beta_{\pm1}$, meaning that all three spin components deplete with the same rate constant, which in turn confirms that the collision ``event rate'' for the $(m'_J,m_J)$ = (0,0) collisions is half that of the (1,-1), since each (0,0) collision results in the loss of two $m_J$ = 0 atoms \cite{Julienne:89,PhysRevA.64.042710}.

Previous measurements of the inelastic loss in unpolarized He$^*$ gases have been performed by measuring ionization rates of samples prepared in magneto-optical traps \cite{PhysRevLett.80.5516,PhysRevLett.82.2848,PhysRevA.60.R761,PhysRevA.61.050702,PhysRevA.73.032713}, with the results of, and discrepancies among, these various measurements summarized in refs. \cite{PhysRevA.73.032713,PhysRevA.64.042710}. These experiments differ from that of the present work in that the losses resulting from interactions between individual spin states are not resolved, and instead an ionization rate, $K^{(unpol)}_{ss}$, is measured for the global decay of the entire unpolarized gas.  To compare with these measurements, our partial two-body loss coefficients can be re-scaled to give the global loss rate using
 \begin{equation}
\beta_{meas} = 18 K^{(unpol)}_{ss} \frac{1}{3} (2\rho_{-1}\rho_{+1} + \rho^2_{0})
\label{eq:globalKss}
\end{equation}
where $\beta_{meas}$ is the measured number loss rate, and $\rho_i$ is the fractional population of the $m_J = i$ substate \cite{PhysRevA.61.050702}.  The numerical factors are defined such that the ionization rate constant $K^{(unpol)}_{ss}$ = $\frac{1}{2}\beta_{meas}$  (i.e. 1 ion detected per 2 lost atoms, see eq. \ref{eq:ionizingprocesses}) for the case of an unpolarized gas where $\rho_{-1}$ = $\rho_{+1}$ = $\rho_0$ = $\frac{1}{3}$.  Our measured $\beta_{00}$ with $\rho_0 = 1$ ($\rho_{\pm1}$ = 0) gives $K^{(unpol)}_{ss}$ = $\frac{1}{6}\beta_{00}$ = 1.10(7) $\times$ 10$^{-10}$ cm$^3$/s. Similarly, if $\rho_{-1}$ = $\rho_{+1}$ = $\frac{1}{2}$, $K^{(unpol)}_{ss}$ = $\frac{1}{3}\frac{\beta_{\pm1}}{2}$ = 1.23(17) $\times$ 10$^{-10}$ cm$^3$/s, where the factor of one half in $\beta_{\pm1}$ reflects the fact that we measure only $N_1$ or $N_{-1}$ to determine $\beta_{\pm1}$, whereas when both states are counted, the rate constant is reduced by half. As before, a systematic uncertainty of $25\%$ applies to these numbers.  These values of $K^{(unpol)}_{ss}$ are in good agreement with those measured in refs. \cite{PhysRevA.60.R761,PhysRevA.73.032713} and are consistent with theoretical predictions \cite{Julienne:89,PhysRevA.61.060703,PhysRevA.64.042710}.

In conclusion, we have achieved Bose-Einstein condensation of $^4$He$^*$ in an optical dipole trap.  We observe efficient evaporative cooling and a long trap lifetime, both of which point favorably towards the possibility of future all-optical production of these condensates as well as the implementation of more complex trapping geometries.  Using this trap, we have also made the first spin state-resolved measurements of two-body loss in a thermal $^4$He$^*$ gas, and find that it is dominated by collisions between ($m_J, m'_J$) = (0,0) and (1,-1) atoms.  We find loss rate coefficients for these individual processes that are in line with theory and compatible with previous measurements of the global loss rate from magneto-optical traps.  Finally, we conjecture that although these relatively high loss rates may prevent the cooling of some non-polarized mixtures to degeneracy, spin mixtures can be created starting from a cold polarized gas whose lifetime will not preclude all experiments.  This is especially true for the case of highly polarized spin mixtures consisting of a small number of $m_J$ = 0 atoms immersed in a gas of $m_J$ = 1 or -1 atoms.


%
%


\begin{acknowledgments}
This work was supported by the French ANR, the IFRAF institute, and
the Euroquam Project CIGMA.  GP is supported by a European Union Marie Curie IIF Fellowship.
\end{acknowledgments}

During the preparation of this paper we became aware of the successful operation of an optical dipole trap for He$^*$ at the VU University in Amsterdam \footnote{W. Vassen, private communication.}



\end{document}